\documentclass[twocolumn,preprintnumbers,amsmath,amssymb]{revtex4}

\usepackage{graphicx}
\usepackage{dcolumn}
\usepackage{bm}

\usepackage{amsmath}
\usepackage{amsfonts}
\usepackage{amssymb}

\newcommand{\beq}{\begin{equation}}
\newcommand{\ber}{\begin{eqnarray}}
\newcommand{\eeq}{\end{equation}}
\newcommand{\eer}{\end{eqnarray}}

\begin{document}

\preprint{SUNY BING 10/24/13}

\title{Parastatistical Factors for Cascade Emission of a Pair of Paraparticles}
\author{Charles A. Nelson}
\email{cnelson@binghamton.edu}

\author{Margarita Kraynova}
\author{Calvin S. Mera}
\author{Alanna M. Shapiro}
  
\affiliation{%
Department of Physics, State University of New York at Binghamton, Binghamton, New York 13902-6016\\
}%


\begin{abstract}
The empirical absence to date of particles obeying parastatistics in high energy collider experiments might be due to their large masses and lack of gauge couplings.
If there is a portal to such particles, they might be cascade emitted as a pair of para-Majorana neutrinos $A_1 \rightarrow A_2 \breve{\nu}_{\alpha}  \breve{\nu}_{\beta}$ or as a pair of scalar paraparticles such as in $A_1 \rightarrow A_2 \breve{A}  \breve{B}$.  In this paper, for an assumed portal Lagrangian, the associated parastatistical factors are obtained for the case of order $p=2$ parastatistics and the, in general differing factors, for the cases of emission of a non-degenerate or a degenerate pair of particles which obey normal statistics.  

\end{abstract}

\maketitle

\section{Introduction}

In the standard model all particles are either fermions or bosons which correspond to order $p=1$ parastatistics.   Parastatistics [1-7] is a generalized statistics associated with the permutation group and is allowed in local relativistic quantum field theory.   Particles obeying parastatistics would be pair produced, the lightest such particles are stable and might be responsible for dark matter and/or dark energy.    If there is a portal to such particles, at a high energy collider these particles might be emitted in a cascade process as a pair of para-Majorana neutrinos $A_1 \rightarrow A_2 \breve{\nu}_{\alpha}  \breve{\nu}_{\beta}$ or as a pair of scalar paraparticles such as in $A_1 \rightarrow A_2 \breve{A}  \breve{B}$.  The paraparticles are denoted by a ``soft" or ``breve" accent.   The statistical factors are calculated for these two pair emission cascades because of  their final $ A_{2} $ or $ B_{2}  $,  versus the empirical difficulties for investigating a cascade to an almost massless final neutrino as in $A_1 \rightarrow \nu_2 \breve{A}  \breve{\nu}$.

In this paper the assumed Lagrangian densities for the cascade processes involve a Majorana spin $1/2$ field
 $\xi$ and a neutral complex scalar field $\mathcal A$ which respectively obey fermi and bose statistics, and also
 their counterparts which obey order p=2 parastatistics $ \breve{\xi}$ and  $ \breve{\mathcal A}$.  We consider this complex $\mathcal A$ field in the particle antiparticle basis with respective corresponding quanta $A$ and $B$.  Similarly, $\breve A$ and $\breve B$ are the quanta for $ \breve{\mathcal A}$. We are assuming there are two $ A_{1,2} $ with $ B_{1,2}  $ mass multiplets with $ m_1 > m_2>>0 $ to kinematically allow these cascade processes, and that, if not for the portal couplings, the paraparticles would only interact gravitationally. The parastatistical factors for these cascade emission processes are calculated and compared with the analogous factors in the case of the emitted pair obeying ordinary statistics and in the case when there is a degeneracy, for instance $A_1 \rightarrow A_2 {\nu}_{a,\alpha}  {\nu}_{a,\beta}$ where there are two kinds of emitted ${\nu}_{a, \alpha}  {\nu}_{a, \beta}$.  The portal Lagrangian densities considered  for these two cases are analogous to those for the para case.  
  
Section II  contains the Lagrangian densities assumed for these cascade processes.  It continues with the calculation to lowest perturbative order of the statistical factors in the case of  $p=2$ parastatistics and in the cases of emission a non-degenerate or a degenerate pair obeying normal statistics.  Section III discusses the different predictions of these three cases.  The tri-linear relations for $p=2$ parastatistics are listed in an appendix.


\section{Cascade Processes with Emission of a Pair of Paraparticles}

\subsection{\label{sec:level2}Lagrangian densities}

For each of the interaction Lagrangian densities, there is the associated normalization the coupling constant.  While the definitions made below are usual normalizations associated with the identity of the fields in normal statistics and in parastatistics, these definitions are arbitrary.   However, these definitions are fixed and are used for each of the cascade processes in the calculation of their associated $c_p$ and $c_d$ statistical factors.  From the values obtained for these factors,  the consequences of alternate normalizations can be easily considered.  
  
Among the usual $p=1$ fields, we consider interactions as in the supersymmetric Wess-Zumino model  [8], but with unrelated coupling constants, so the interaction
densities involving only $p=1$ fields are
\ber
{\mathcal L}_{\mathcal Y}= - \frac{f}{2}  (\mathcal{A} \xi \xi +{\mathcal A}^{\dag} \bar{\xi} \bar{\xi})
\eer
\ber
{\mathcal L}_{C}= - \frac{t}{2} { \{ \mathcal A} ({\mathcal A}^{\dag})^{2 } +{\mathcal A}^2 {\mathcal A}^{\dag}\} 
\eer
\ber
{\mathcal L}_{q}= - \frac{F}{4} ({\mathcal A}^{\dag})^{2 }  {\mathcal A}^2  
\eer

For the cascade processes, we consider the following portal couplings between these p=1 fields and the order p=2 fields, with anticommutator curly braces and commutator square brackets:
\ber
{\mathcal L}_{\breve{\mathcal Y} }= -  \frac{\breve f}{2}([\breve{\xi}, \breve{\xi}] \mathcal A 
+ {\mathcal A}^{\dag} [\bar{\breve{\xi}} ,\bar{\breve{\xi}}] )
\eer
\ber
{\mathcal L}_{2 \breve{c}}= - \frac{\breve{t}}{2} (  \{ { \breve{\mathcal A}} , {\breve{\mathcal A}} \}  {\mathcal A}^{\dag}  
+ {\mathcal A}  \{ {\breve{\mathcal A}}^{\dag} , {\breve{\mathcal A}}^{\dag} \} )
\eer
\ber
{\mathcal L}_{3 \breve{c}}= - \frac{\breve{T}}{2} ( {\mathcal A} +  {\mathcal A}^{\dag} )
 \{ {\breve{\mathcal A}} , {\breve{\mathcal A}}^{\dag} \} )
\eer
\ber
{\mathcal L}_{2 \breve{q}}= - \frac{\breve{F}}{4} (  \{ { \breve{\mathcal A}} , {\breve{\mathcal A}} \} ( {\mathcal A}^{\dag} )^2 
+  {\mathcal A} ^2  \{ {\breve{\mathcal A}}^{\dag} , {\breve{\mathcal A}}^{\dag} \} )
\eer
\ber
{\mathcal L}_{3 \breve{q}}= - \frac{\breve{G}}{4} \{ { {\mathcal A}} , {\mathcal A}^{\dag} \}
\{ {\breve{\mathcal A}} , {\breve{\mathcal A}}^{\dag} \}
\eer
\ber
{\mathcal L}_{\breve{\mathcal A} }= -  \frac{\breve j}{2}( \xi \{ \breve{\xi} , \breve{ \mathcal A } \}
+ \{  {\breve {\mathcal A}}^{\dag} , \bar{\breve{\xi}} \}   \bar{\xi} )
\eer

For comparison, instead of fields obeying $p=2$ parastatistics, we also consider the case with pair emission fields ${\mathcal A}_a$  and $\xi_a$ 
obeying bose and fermi statistics.   For a non-degenerate pair, this $a$ subscript is single-valued.  It is two-valued and will be summed over in emission of a degenerate pair.   These Lagrangian densities are analogous to the above portal ones:
\ber
{\mathcal L}^d_{\mathcal Y}= - \frac{f_d}{2}  (\mathcal{A} \xi_a \xi_a +{\mathcal A}^{\dag} \bar{\xi}_a \bar{\xi}_a)
\eer
\ber
{\mathcal L}^d_{2c}= -  \frac{t_d}{2} \{ {\mathcal A}^{\dag}  {\mathcal A}_a {\mathcal A}_a  
+ {\mathcal A}^{\dag}_a {\mathcal A}^{\dag}_a {\mathcal A} \} 
\eer
\ber
{\mathcal L}^d_{3c}= - {T_d}  \{ {\mathcal A}  {\mathcal A}_a {\mathcal A}^{\dag}_a  
+{\mathcal A}_a {\mathcal A}^{\dag}_a {\mathcal A}^{\dag} \} 
\eer
\ber
{\mathcal L}^{d}_{2q}= - \frac{J_d}{2}  ( {\mathcal A}_a {\mathcal A}_a )  ( {\mathcal A}^{\dag}_b {\mathcal A}^{\dag}_b )
\eer
\ber
{\mathcal L}^{d}_{3q}= - G_{d} ( {\mathcal A}_a {\mathcal A}^{\dag}_a )  ( {\mathcal A} {\mathcal A}^{\dag} )
\eer 
\ber
{\mathcal L}^d_{\mathcal A}= - {j_d}  ( \xi \xi_a \mathcal{A}_a +{\mathcal A}^{\dag}_a \bar{\xi}_a \bar{\xi})
\eer

\subsection{\label{sec:level2}Parastatistical factors for cascade processes}

The above interaction Lagrangian densities have a particle-antiparticle transformation symmetry such that the results obtained for each cascade also hold for the cascade obtained by transforming all $ {A}_i \leftrightarrow   {B}_i $  and $ \breve{A}_j \leftrightarrow   \breve{B_j}$, for instance the parastatistical factors are the same for $A_1 \rightarrow A_2 \breve{A}  \breve{B}$ and $B_1 \rightarrow B_2 \breve{B}  \breve{A}$.    For the comparison normal statistics cases involving ${\mathcal A}_a$  and $\xi_a$, there is the analogous transformation of all $ {A}_i \leftrightarrow   {B}_i $  and $A_{a,j} \leftrightarrow   B_{a,j}$.  


\subsubsection{Emission of a pair of para-Majorana neutrinos}

In evaluation of the S-matrix elements for the cascade processes, we evaluate amplitudes in the occupation number basis for the paraparticles in the final state and then construct the corresponding amplitudes in the permutation group basis for the physical paraparticles.  We omit disconnected diagrams and ones with ${\mathcal L}_{int}$ self-contractions, that is we require each field in ${\mathcal L}_{int}$ contract with a field in a different ${\mathcal L}_{int}$ or with a particle in the initial or final state.  
For a cascade by emission of a pair of para-Majorana neutrinos $A_1 \rightarrow A_2 \breve{\nu}_{\alpha}  \breve{\nu}_{\beta}$, for 
${\mathcal L}_{\breve{\mathcal Y} }$ and ${\mathcal L}_{C}$ there is the time-ordered  
\ber
S_{fi}= (i \frac{t}{2} )( i \frac{{\breve{f}} }{2} )  \int d^4 x_1  \int d^4 x_2 ~ 
\theta( t_1 - t_2) \nonumber \\
{}_A\!\!< A_l \breve{\nu}_{\alpha}  \breve{\nu}_{\beta}| \{ 
{\mathcal L}_{\breve{\mathcal Y} }(x_1)    {\mathcal L}_{C}(x_2)     +      {\mathcal L}_{C}(x_1) {\mathcal L}_{\breve{\mathcal Y} }(x_2)
\} |A_k> \nonumber \\
\eer
The final state has the $ \breve{\nu}_{\alpha}  \breve{\nu}_{\beta} $ operators in the A-order 
$ | A_l  \breve{\nu}_{\alpha}  \breve{\nu}_{\beta} >_A = \frac{1}{2} l^{\dag} \alpha^{\dag} \beta^{\dag} |0> $ in the occupation number basis. 
 For the B-ordered state, the order for the two paraparticles is reversed $ | A_l  \breve{\nu}_{\alpha}  \breve{\nu}_{\beta} >_B = \frac{1}{2} l^{\dag}  \beta^{\dag}\alpha^{\dag} |0> $.   Here we have suppressed the covariant normalization factors $ (2\pi)^{ \frac{3}{2} } \sqrt{2 E}$ for each particle in the external states.
 
For the A-ordered final state, by writing the fields in their positive and negative frequency parts and then using the $p=2$ tri-linear relations for the paraquanta, we obtain amplitudes corresponding to the two diagrams in Fig. 1.  
\begin{figure}
\includegraphics[ trim= 420 0 0 250, scale=0.55 ]{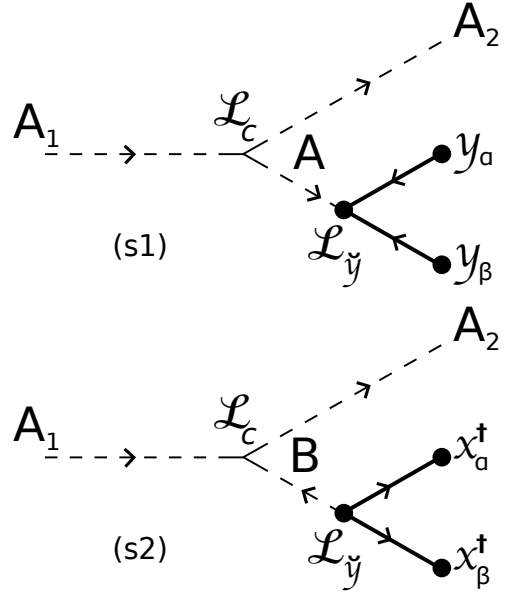}
\caption{\label{fig:epsart} First 2 diagrams for a cascade by emission of a pair of para-Majorana neutrinos $A_1 \rightarrow A_2 \breve{\nu}_{\alpha}  \breve{\nu}_{\beta}$. $B$ denotes the antiparticle to $A$. }
\end{figure}
To maintain simplicity of the expressions for the matrix elements, we omit the associated mixing matrices between the mass eigenstates and the interaction eigenstates.   From (16), the (s1) amplitude for the A-ordered final state is
\ber
- i \mathcal{M}^{(s1)}_A = \{  1 \} (i t)( i \breve{f} ) \frac{i}{q^2_A -{m^2} + i \epsilon}  y ( \vec{\alpha},\lambda_{\alpha} )^A 
y ( \vec{\beta},\lambda_{\beta} )_A  \nonumber \\
\eer
with A the 2-valued index for the 2-component spinor [9], and the (s2) amplitude is
\ber
- i \mathcal{M}^{(s2)}_A = \{  1 \} (i t)( i \breve{f} ) \frac{i}{q^2_B -{m^2} + i \epsilon}  x^{\dag} ( \vec{\alpha},\lambda_{\alpha} )_{\dot{A}} 
x^{\dag} ( \vec{\beta},\lambda_{\beta} )^{\dot{A}} 
 \nonumber \\
\eer

 In the case with pair emission fields ${\mathcal A}_a$  and $\xi_a$  obeying bose and fermi statistics, the same amplitudes for (s1) and (s2) are obtained
for the process $A_1 \rightarrow A_2 {\nu}_{a,\alpha}  {\nu}_{a,\beta}$ with $ | A_l  {\nu}_{a,\alpha}  {\nu}_{a,\beta} > =  l^{\dag} \alpha_a^{\dag} \beta_a^{\dag} |0> $ except in place of the parastatistics factor $\{  1 \} ( t \breve{f} )$ there is the factor $\{  1 \} (t  f_d )$, where the respective statistical factors ${c_p} $ and ${c_d}$ are given in the curly braces.   In writing these statistics factors times coupling constants, we omit each $( i)$  associated with the $ i {\cal{L}}_{int}$ vertex. This is the comparison amplitude for all fields obeying ordinary statistics for the Lagrangian densities given above.   When there are two kinds of emitted ${\nu}_{a,\alpha}  {\nu}_{a,\beta}$, in calculation of the partial decay width, a factor of 2 appears due summing over the two final degenerate channels.

For the B-ordered final state, the same amplitudes for (s1) and (s2) are obtained but with opposite overall sign versus the A-ordered final state, so the permutation group basis amplitudes $\mathcal{M}^{(s1)} $ and $\mathcal{M}^{(s2)} $ for the symmetric/antisymmetric final states
\ber
 | A_l  \breve{\nu}_{\alpha}  \breve{\nu}_{\beta} >_{sym,asym} = \frac{1}{\sqrt{2}}  ( | A_l  \breve{\nu}_{\alpha}  \breve{\nu}_{\beta} >_A \pm | A_l  \breve{\nu}_{\alpha}  \breve{\nu}_{\beta} >_B ) \nonumber \\
 \eer
  are respectively zero and $\sqrt{2}$ times those for the A-ordering.    Hence, from the values of the statistical factors ${c_p} $ and ${c_d}$,  if these were the only two diagrams, upon summing over the two permutation basis final states for the decay process $A_1 \rightarrow A_2 \breve{\nu}_{\alpha}  \breve{\nu}_{\beta}$ the rate would be twice that for for the corresponding normal statistics process $A_1 \rightarrow A_2 {\nu}_{a,\alpha}  {\nu}_{a,\beta}$ with a non-degenerate pair, but the $p=2$ rate would be the same as that for the case of emission of two kinds of ${\nu}_{a,\alpha}  {\nu}_{a,\beta}$ due to summing over these two degenerate channels. 

For $A_1 \rightarrow A_2 \breve{\nu}_{\alpha}  \breve{\nu}_{\beta}$ there is also a contribution to second order in   
${\mathcal L}_{\breve{\mathcal Y} }$ which corresponds to the two diagrams in Fig. 2. 
\begin{figure}
\includegraphics[ trim= 420 0 0 250, scale=0.55 ]{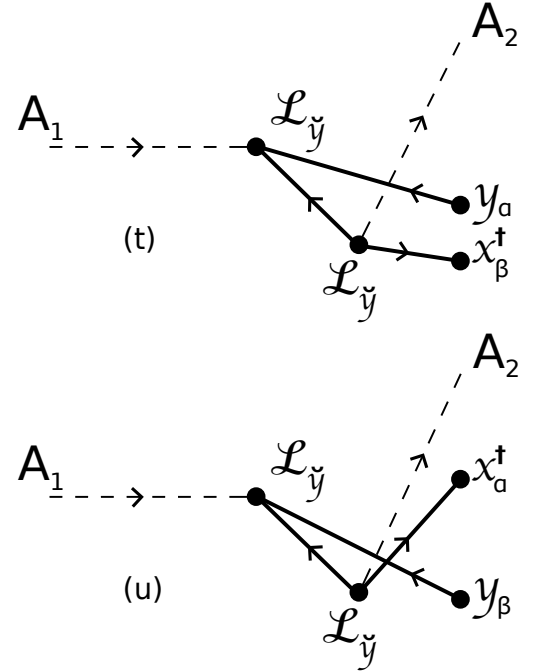}
\caption{\label{fig:epsart} The remaining 2 diagrams for $A_1 \rightarrow A_2 \breve{\nu}_{\alpha}  \breve{\nu}_{\beta}$.}
\end{figure}
Again, for each diagram, the B-ordering gives the same amplitude but with opposite overall sign versus the A-ordering.  Also, again for  the A-ordering the expressions associated with the diagrams are proportional in the case of paraparticles and the case of non-degenerate Majorana fermions. The contribution of the $(u)$ diagram is minus that of the $(t)$ diagram with $\alpha \leftrightarrow \beta$ exchanged.  In the para case, the $(t)$ diagram has a factor $\{  4 \} ( \breve{f} )^2 $ and in the fermion case there is instead $\{  \frac{1}{2} \} (f_d )^2$, so the respective statistical factors ${c_p} $ and ${c_d}$ now differ, unlike for the previous $(s1)$ and $(s2)$ diagrams.  

As shown in Fig. 3, there is a similar cascade from $A_1$ to the antiparticle $B_2$ by emission of a pair of para-Majorana neutrinos, $A_1 \rightarrow B_2 \breve{\nu}_{\alpha}  \breve{\nu}_{\beta}$:
\begin{figure}
\includegraphics[ trim= 420 0 0 250, scale=0.55 ]{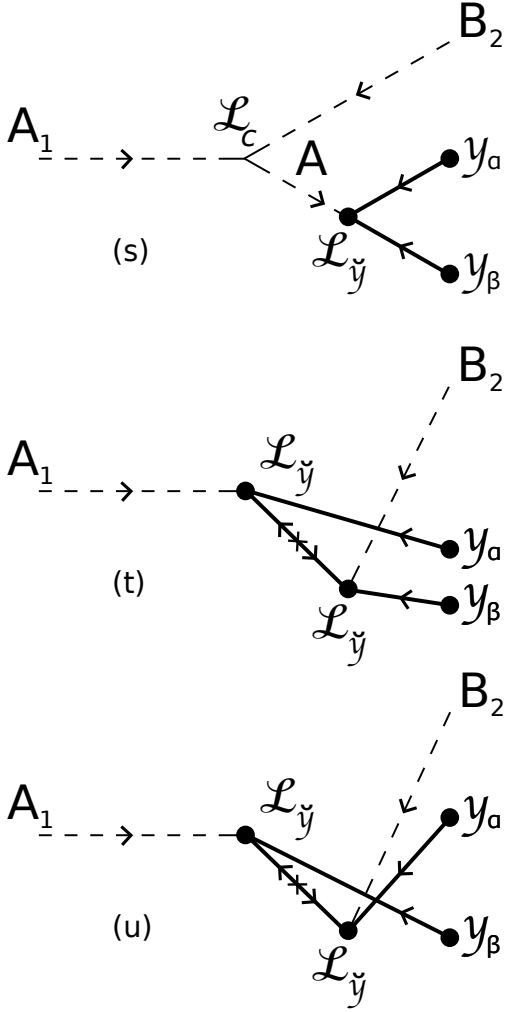}
\caption{\label{fig:epsart}  The 3 diagrams for cascade to $B_2$ by emission of a pair of para-Majorana neutrinos $A_1 \rightarrow B_2 \breve{\nu}_{\alpha}  \breve{\nu}_{\beta}$.  }
\end{figure} 
For each diagram, for the A-ordering the para amplitude is proportional to that in the case of Majorana fermions.  Also for each diagram, the B-ordered expression is of opposite sign to that for the A-ordering, so the permutation group basis amplitude is again the asymmetric one as for the previous process $A_1 \rightarrow A_2 \breve{\nu}_{\alpha}  \breve{\nu}_{\beta}$ but the diagrams are different for $A_1 \rightarrow B_2 \breve{\nu}_{\alpha}  \breve{\nu}_{\beta}$ with the final $B_2$:  From ${\mathcal L}_{\breve{\mathcal Y} }$ and ${\mathcal L}_{C}$, for the A-ordering there is a single $(s)$ diagram with the parastatistics factor $\{  1 \} ( t \breve{f} )$.    For the analogous all $p=1$ cascade $A_1 \rightarrow B_2 {\nu}_{a,\alpha}  {\nu}_{a,\beta}$, there is the factor $\{  1 \} (t  f_d )$.  The second order contribution in ${\mathcal L}_{\breve{\mathcal Y} }$ involves a Majorana mass insertion contribution, and the contribution of the $(u)$ diagram is again  negative that for the $(t)$ diagram with $\alpha \leftrightarrow \beta$ exchanged.  For the $(t)$ diagram, in the para case in the there is the factor   $\{  4 \} ( \breve{f} )^2 $ and correspondingly in the fermion case $\{  \frac{1}{2} \} (f_d )^2$.  

For the cascade processes considered in this paper, this value of 8  is the largest for the ratio   ${c_p}/{c_d}$ for the associated amplitudes.  It could give a strong test between the paraparticle and both fermion pair emission cases if the contribution of the $(t)$ and $(u)$ diagrams were to dominate for $A_1 \rightarrow A_2 \breve{\nu}_{\alpha}  \breve{\nu}_{\beta}$ and/or $A_1 \rightarrow B_2 \breve{\nu}_{\alpha}  \breve{\nu}_{\beta}$
in some kinematic region.


\subsubsection{Emission of a pair of scalar paraparticles}

In the remaining five cascade processes, a pair of scalar paraparticles are emitted.  For each process, the obtained A-ordered amplitudes can again be considered in terms of diagrams as shown in the figures.  These A-amplitudes in the para case are again proportional to those in the non-degenerate case in which there is a boson scalar pair emitted.  In the following, for each diagram the respective statistical factors ${c_p} $ and ${c_d}$ are listed. 

For these processes with emission of a pair of scalar particles, for the B-ordered final state, the same amplitudes are obtained as for the A-ordered final state, so in all cases in the permutation group basis, the associated symmetric final state has an amplitude $\sqrt{2}$ times that for the A-ordering, and the amplitude for the antisymmetric final state vanishes.  For instance, for the first process $A_1 \rightarrow A_2 \breve{A}  \breve{B}$ with emission of a particle-antiparticle pair of paraparticles, the symmetric/antisymmetric final states are 
\ber
 | A_{2,l} \breve{A}  \breve{B}  >_{sym,asym} = \frac{1}{\sqrt{2}}  ( | A_{2,l}  \breve{A}  \breve{B} >_A \pm | A_{2,l}  \breve{A}  \breve{B}>_B ) \nonumber \\
 \eer
with A-ordering $ | A_{2,l} \breve{A}  \breve{B} >_A = \frac{1}{2} l^{\dag} A^{\dag} B^{\dag} |0> $ and B-ordering $ | A_{2,l}  \breve{A}  \breve{B} >_B = \frac{1}{2} l^{\dag} B^{\dag} A^{\dag} |0> $.   
In the case of boson pair emission in the corresponding process $A_1 \rightarrow A_2 A_a B_a $, the final state is 
$ | A_{2,l}  A_a B_a > =  l^{\dag} A_a^{\dag} B_a^{\dag} |0> $.  For a degenerate bosonic pair emitted, there would be a factor of 2 in the rate, so when $c_p = c_d$ if there were only that diagram contributing, there is the same rate in the case of of paraparticle emission $\breve{A}  \breve{B}$ upon summing over the two permutation group basis final states and in the case of degenerate pair emission of two kinds of $A_a B_a$.

Fig. 4 shows the first 3 diagrams for the cascade $A_1 \rightarrow A_2 \breve{A}  \breve{B}$.
\begin{figure}
\includegraphics[ trim= 420 0 0 250, scale=0.55 ]{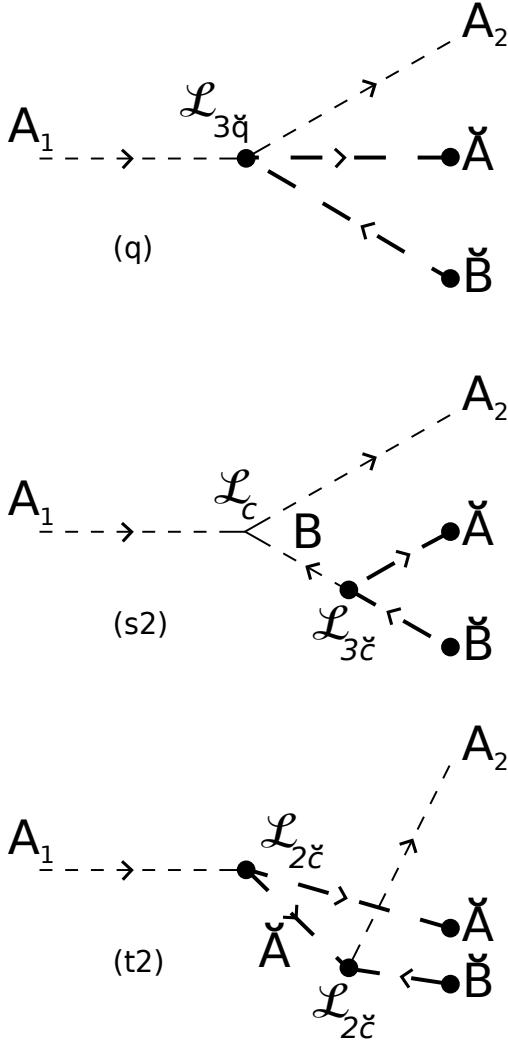}
\caption{\label{fig:epsart} First 3 diagrams for the cascade $A_1 \rightarrow A_2 \breve{A}  \breve{B}$ by emission of a particle-antiparticle pair of scalar paraparticles $\breve{A} \breve{B}$.  $\breve{B}$ denotes the antiparticle to $\breve{A}$.}
\end{figure}
Fig. 5 shows the remaining 3 diagrams: 
\begin{figure}
\includegraphics[ trim= 420 0 0 250, scale=0.55 ]{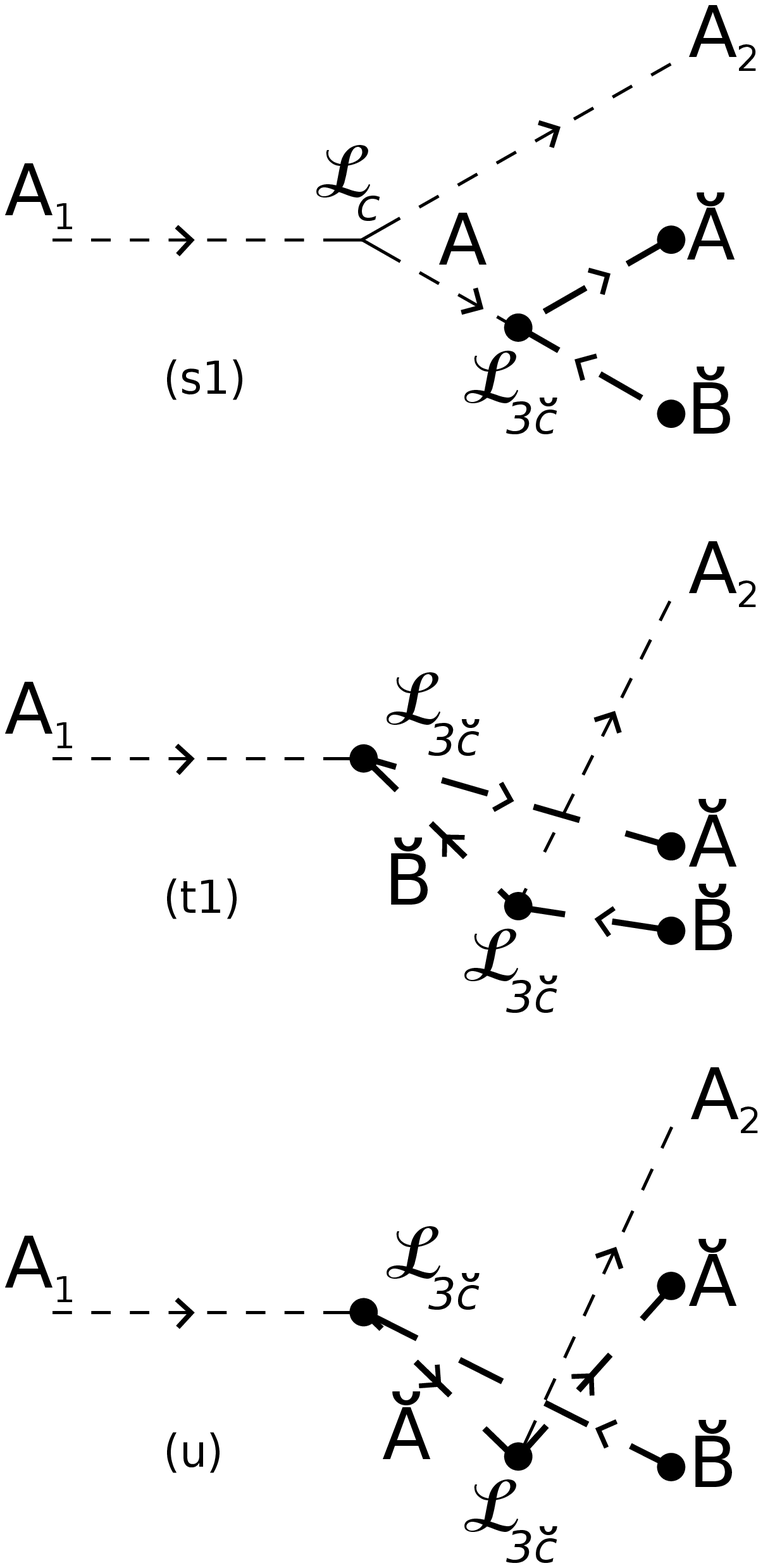}
\caption{\label{fig:epsart} The remaining 3 diagrams for $A_1 \rightarrow A_2 \breve{A}  \breve{B}$.}
\end{figure}
From ${\mathcal L}_{3 \breve{q}}$, there is  the $(q)$ diagram with the factors
$\{  1 \} ( - \breve{G} )$ versus $\{  1 \} ( - G_d )$. The minuses occur here because we omit each $( i)$  associated with the $ i {\cal{L}}_{int}$ vertex.  From $ {\mathcal L}_{C} $ and ${\mathcal L}_{3 \breve{c}} $,  the $(s1)$ and $(s2)$ diagrams each have the factors $\{1 \} ( t \breve{T} )$ versus $\{  1 \} ( t T_d )$.  From second order in ${\mathcal L}_{3 \breve{c}} $,  the $(t1)$ and $(u)$ diagrams each have the factors $\{  \frac{1}{2} \} ( \breve{T} )^2$ versus $\{  1 \} ( T_d )^2$.  
There is only a single diagram contribution for second order ${\mathcal L}_{2 \breve{c}} $.  This $(t2)$ diagram has the factors $\{  2 \} ( \breve{t} )^2$ versus $\{  1 \} ( t_d )^2$.

The analogous cascade from $A_1$ to the antiparticle $B_2$, $A_1 \rightarrow B_2 \breve{A}_3 \breve{A}_4 $, has the 6 diagrams shown in Figs. 6 and 7:
\begin{figure}
\includegraphics[ trim= 420 0 0 250, scale=0.55 ]{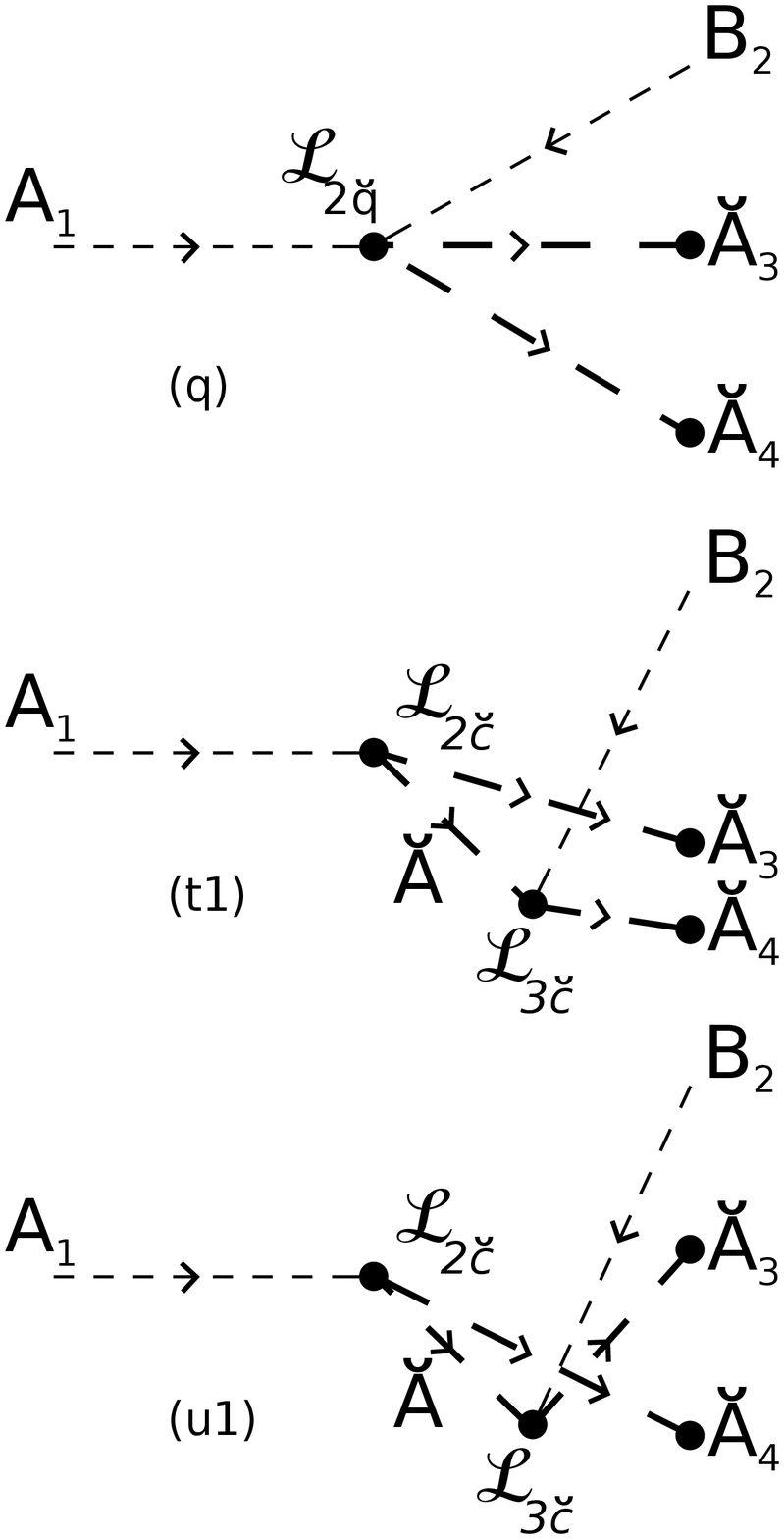}
\caption{\label{fig:epsart} First 3 diagrams for the cascade $A_1 \rightarrow B_2 \breve{A}_3 \breve{A}_4 $ by emission of a pair of scalar paraparticles $\breve{A}_3 \breve{A}_4$.}  
\end{figure}
\begin{figure}
\includegraphics[ trim= 420 0 0 250, scale=0.55 ]{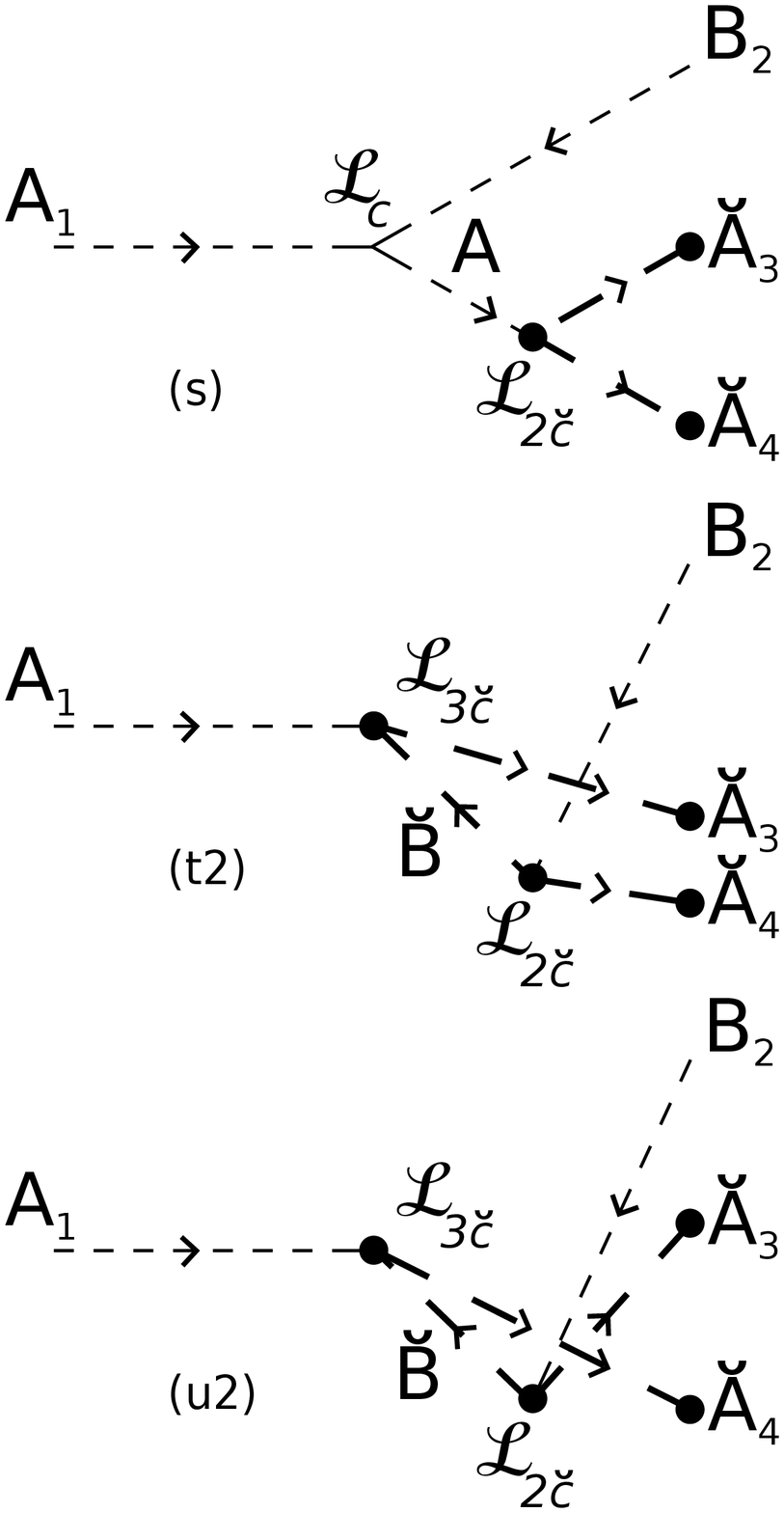}
\caption{\label{fig:epsart} The remaining 3 diagrams for $A_1 \rightarrow B_2 \breve{A}_3 \breve{A}_4 $.} 
\end{figure}

From ${\mathcal L}_{2 \breve{q}}$, there is the $(q)$ diagram with the factors
$\{  1 \} ( -  \breve{F} )$ in the para case versus factors $\{  1 \} ( -  F_d )$ in the boson case. From $ {\mathcal L}_{C} $ and ${\mathcal L}_{2 \breve{c}} $,  the $(s)$ diagram has the factors $\{2 \} ( t \breve{t} )$ versus $\{  1 \} ( t t_d )$.  As shown, the remaining four diagrams arise from $ {\mathcal L}_{2 \breve{c}}$ and ${\mathcal L}_{3 \breve{c}} $.  Each of $(t1)$, $(u1)$, $(t2)$ and $(u2)$ has the factors   $\{  1 \} ( \breve{t} \breve{T} )   $ versus $ \{  1 \} ( t_d  T_d ) $.

The cascade from $A_1$ to $A_2$ by $A_1 \rightarrow A_2 \breve{A}_3 \breve{A}_4 $  has the diagrams shown in Fig.8:  
\begin{figure}
\includegraphics[ trim= 420 0 0 250, scale=0.55 ]{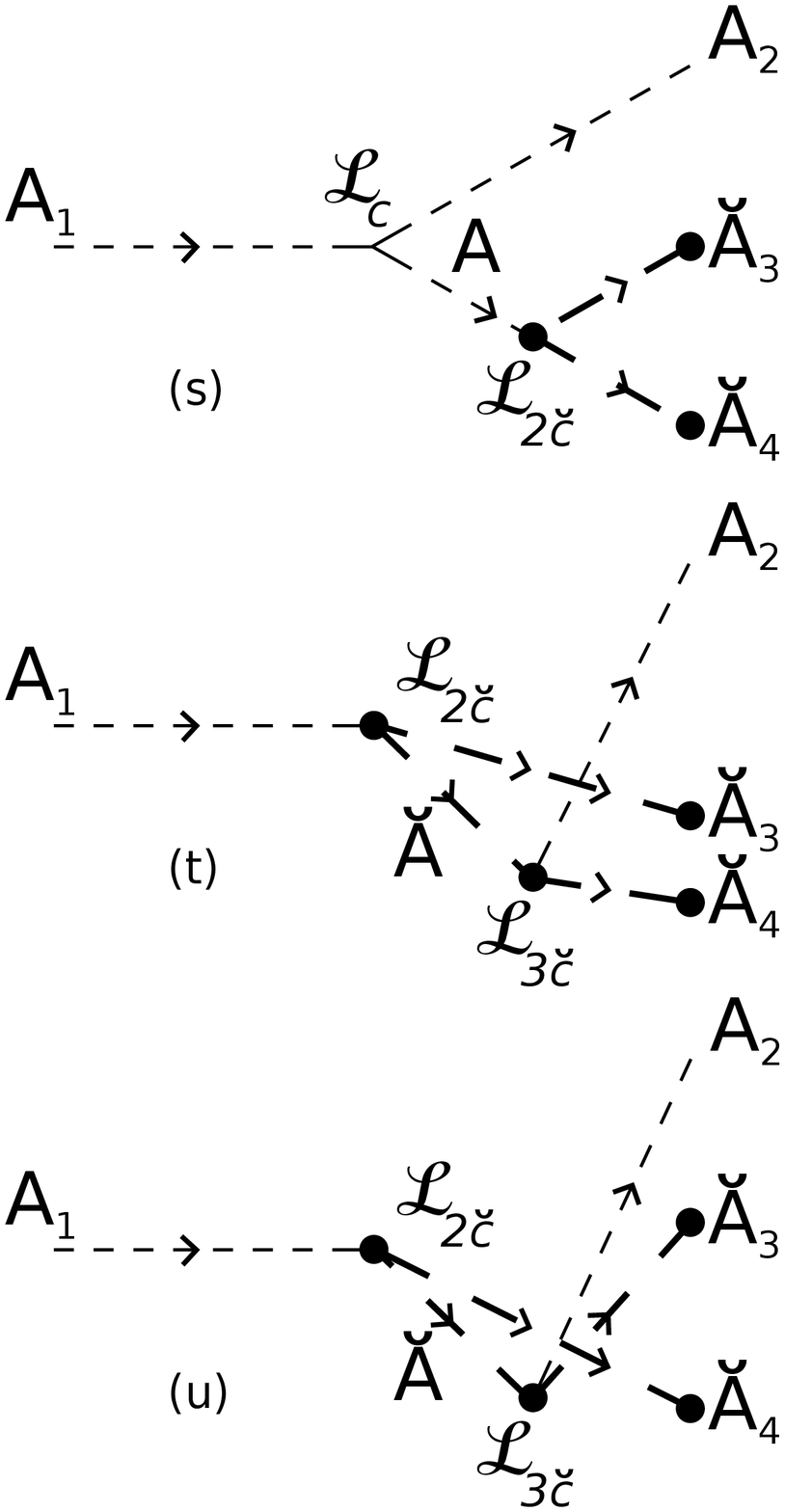}
\caption{\label{fig:epsart} 
The 3 diagrams for the cascade $A_1 \rightarrow A_2 \breve{A}_3 \breve{A}_4 $ by emission of a pair of scalar paraparticles $\breve{A}_3 \breve{A}_4$.} 
\end{figure}
From ${\mathcal L}_{C} $ and ${\mathcal L}_{2 \breve{c}} $,  the $(s)$ diagram has the factors $\{2 \} ( t \breve{t} )$ versus $\{  1 \} ( t t_d )$.  From $ {\mathcal L}_{2 \breve{c}} $ and ${\mathcal L}_{3 \breve{c}} $,  the $(t)$ and $(u)$ diagrams each have the factors $\{1 \} ( \breve{t} \breve{T} )$ versus $\{  1 \} ( t_d T_d )$.

If instead there is emission of antiparticle pair $\breve{B}_3 \breve{B}_4$ via the cascade $A_1 \rightarrow A_2 \breve{B}_3 \breve{B}_4 $,
there are the diagrams shown in Fig. 9:  
\begin{figure}
\includegraphics[ trim= 420 0 0 250, scale=0.55 ]{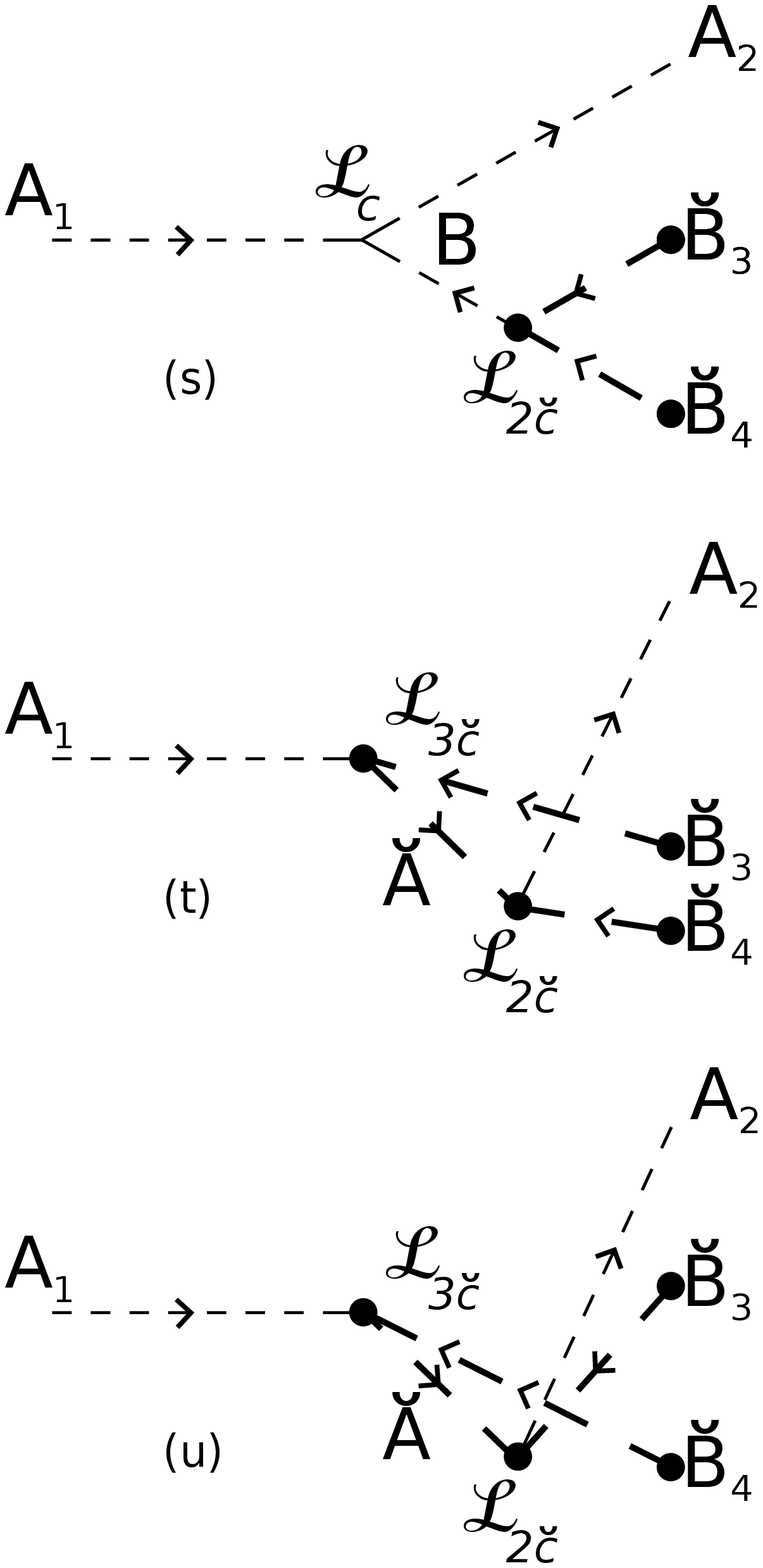}
\caption{\label{fig:epsart} The 3 diagrams for the cascade $A_1 \rightarrow A_2 \breve{B}_3 \breve{B}_4 $ by emission of an antiparticle pair of scalar paraparticles $\breve{B}_3 \breve{B}_4$.}
\end{figure}

From ${\mathcal L}_{C} $ and ${\mathcal L}_{2 \breve{c}} $,  the $(s)$ diagram has the factors $\{2 \} ( t \breve{t} )$ versus $\{  1 \} ( t t_d )$.  From  $ {\mathcal L}_{2 \breve{c}} $ and ${\mathcal L}_{3 \breve{c}} $,  the $(t)$ and $(u)$ diagrams each have the factors $\{1 \} ( \breve{t} \breve{T} )$ versus $\{  1 \} ( t_d T_d )$.  The  interaction $ {\mathcal L}_{2 \breve{c}} $ and ${\mathcal L}_{3 \breve{c}} $ vertices in the $(t)$ and $(u)$ diagrams are exchanged in Fig. 9 for emission of $\breve{B}_3 \breve{B}_4$ versus those in Fig. 8 for emission of $\breve{A}_3 \breve{A}_4$. 

For the cascade $A_1 \rightarrow B_2 \breve{A}_3 \breve{B}_4 $ by emission of $\breve{A}_3 \breve{B}_4$ there are the diagrams in Fig. 10:

From ${\mathcal L}_{C} $ and ${\mathcal L}_{3 \breve{c}} $,  the $(s)$ diagram has the factors $\{1 \} ( t \breve{T} )$ versus $\{  1 \} ( t T_d )$.  From  second order in $ {\mathcal L}_{3 \breve{c}} $,  the $(t)$ and $(u)$ diagrams each have the factors $\{ \frac{1}{2} \} (\breve{T} )^2$ versus $ \{  1 \} ( T_d )^2 $.  
\linebreak


\newpage

\section{Discussion}

In general, for the Lagrangian densities considered above, no single cascade process has the same values for all its diagrams for both the $c_p$ and $c_d$ statistical factors which would enable factorization of these factors into overall coefficients.  Possible redefinitions of some of the coupling constants, so as to achieve this for some of the cascade processes could be useful in consideration of specific processes to empirically compare the para case with the cases of emission of a non-degenerate or a degenerate pair of particles which obey normal statistics.  

In the special case when all the interaction densities involving only the $p=1$ fields are absent, all but one of the cascade processes considered above does have common values for both the $c_p$ and $c_d$ statistical factors.  The exception is the process $A_1 \rightarrow A_2 \breve{A}_3 \breve{B}_4 $.  For the other cascades, overall factorization of these statistical factors occurs and the partial decay rates for the para case versus that for emission of a non-degenerate pair are related by
\ber
d {\Gamma}_{p=2} =   2 ~ | {c_p}/{c_d} |^2 ~  d {\Gamma}_{p=1}
 \eer
where the 2 permutation group final states have been summed in the para case.  This relation assumes that the corresponding coupling constants involved in the cascade are equal in the para and $p=1$ cases.  From this expression,  for cascade emission of a pair of para-Majorana neutrinos the partial decay rate would be enhanced by two orders of magnitude for emission of $ \breve{\nu}_{\alpha}  \breve{\nu}_{\beta} $  versus ${\nu}_{a,\alpha}  {\nu}_{a,\beta} $   due to the values of $c_p$ and $c_d$ obtained in Section II.   

Similarly, when all the interaction densities involving only the $p=1$ fields are absent, the partial rate for the para case can be compared with that for the case of emission of a degenerate pair  
\ber
d {\Gamma}_{p=2} =    | {c_p}/{c_d} |^2 ~  d {\Gamma}_{deg. pair}
 \eer
Versus emission of a degenerate pair obeying normal statistics, there are different predictions for the partial rates for emission of a pair of para-Majorana neutrinos and for $A_1 \rightarrow B_2 \breve{A}_3 \breve{B}_4 $, but the same partial rates are predicted for $A_1 \rightarrow B_2 \breve{A}_3 \breve{A}_4 $, $A_1 \rightarrow A_2 \breve{A}_3 \breve{A}_4 $, and $A_1 \rightarrow A_2 \breve{B}_3 \breve{B}_4 $.

While for the process $A_1 \rightarrow A_2 \breve{A}_3 \breve{B}_4 $ there is not overall factorization of these statistical coefficients, this process does have diagrams with different valued $c_p$ and  $c_d$ coefficients, and the $(q)$ diagram has the same values for these coefficients, so by this cascade process there may be potential tests of the para case versus the cases of emission of a non-degenerate or a degenerate pair of particles which obey normal statistics. 

\newpage


\appendix*
\section{Tri-Linear Relations for $p=2$ Parastatistics }

In the calculations of the cascade matrix elements, the following tri-linear relations for order $p=2$ parastatistics are used with parabose operators denoted with Roman letters and parafermi operators denoted with Greek letters.  The mode index $k,l,m$ includes the momentum components, and the helicity components for the para-Majorana field $\breve{\xi}$, and  the $\breve{A}$, $\breve{B}$ distinction for the $\breve{\mathcal{A}}$ complex field.   As for the usual $p=1$ bi-linear relations, in each relation the left-hand-side has the second term written in opposite order from the first term.  The second term has a plus (minus) sign when mostly parabosons (parafermions) occur in the tri-linear relation.  On the right-hand-side, the existence of a Kronecker delta term, and its sign, corresponds with the $a_k a^{\dag}_l$ or $ {\alpha}_k {\alpha}^{\dag}_l$ from the left-hand-side.  The tri-linear relations maintain the associated odd (even) place positions of the operators, whether reading left-to-right, or right-to-left.  These properties also occur in the adjointed relations.  The normalization of these $p=2$ relations corresponds to that of the trilinear relations for arbitrary $p$ parastatistics.  The usual $p=1$ creation and annihilation operators for $\mathcal{A}$    commute  with these $p=2$ operators and those for $\xi$ commute (anticommute) with the parabosons ( parafermions).  

For all parabosons:  
\ber
a_k a_l a_m - a_m a_l a_k =0,   \nonumber \\
a_k a_l a_m^{\dag} - a_m^{\dag} a_l a_k = 2 \delta_{lm} a_k \nonumber \\
a_k a_l^{\dag} a_m - a_m a_l^{\dag} a_k = 2 \delta_{kl} a_m -2 \delta_{ml} a_k \nonumber \\
\eer
For all parafermions:
\ber
{\alpha}_k {\alpha}_l {\alpha}_m + {\alpha}_m {\alpha}_l {\alpha}_k =0,   \nonumber \\
{\alpha}_k {\alpha}_l {\alpha}_m^{\dag} + {\alpha}_m^{\dag} {\alpha}_l {\alpha}_k = 2 \delta_{lm} {\alpha}_k \nonumber \\
{\alpha}_k {\alpha}_l^{\dag} {\alpha}_m + {\alpha}_m {\alpha}_l^{\dag} {\alpha}_k = 2 \delta_{kl}{\alpha}_m +2 \delta_{ml} {\alpha}_k \nonumber \\
\eer
For two parabosons and one parafermion:
\ber
a_k a_l {\beta}_m - {\beta}_m a_l a_k =0,   \nonumber \\
a_k {\beta}_l a_m - a_m {\beta}_l a_k =0,   \nonumber \\
a_k a_l {\beta}^{\dag}_m - {\beta}^{\dag}_m a_l a_k =0,   \nonumber \\
a_k {\beta}_l a^{\dag}_m - a^{\dag}_m {\beta}_l a_k =0,   \nonumber \\
{\beta}_k a_l a_m^{\dag} - a_m^{\dag} a_l {\beta}_k = 2 \delta_{lm} {\beta}_k \nonumber \\
a_k a_l^{\dag} {\beta}_m - {\beta}_m a_l^{\dag} a_k = 2 \delta_{kl} {\beta}_m \nonumber \\
a_k {\beta}_l^{\dag} a_m - a_m {\beta}_l^{\dag} a_k = 0 \nonumber \\
\eer

For two parafermions and one paraboson:
\ber{\alpha}_k {\alpha}_l b_m + b_m {\alpha}_l {\alpha}_k =0,   \nonumber \\
{\alpha}_k b_l {\alpha}_m + {\alpha}_m b_l {\alpha}_k =0,   \nonumber \\
{\alpha}_k {\alpha}_l b^{\dag}_m + b^{\dag}_m {\alpha}_l {\alpha}_k =0,   \nonumber \\
{\alpha}_k b_l {\alpha}^{\dag}_m + {\alpha}^{\dag}_m b_l {\alpha}_k =0,   \nonumber \\
b_k {\alpha}_l {\alpha}^{\dag}_m+ {\alpha}^{\dag}_m {\alpha}_l b_k = 2 \delta_{lm} b_k \nonumber \\
{\alpha}_k {\alpha}^{\dag}_l b_m + b_m {\alpha}^{\dag}_l {\alpha}_k = 2 \delta_{kl} {b}_m \nonumber \\
{\alpha}_k b_l^{\dag} {\alpha}_m + {\alpha}_m b_l^{\dag} {\alpha}_k = 0 \nonumber \\
\eer

\newpage
\begin{figure}
\includegraphics[ trim= 420 0 0 250, scale=0.55 ]{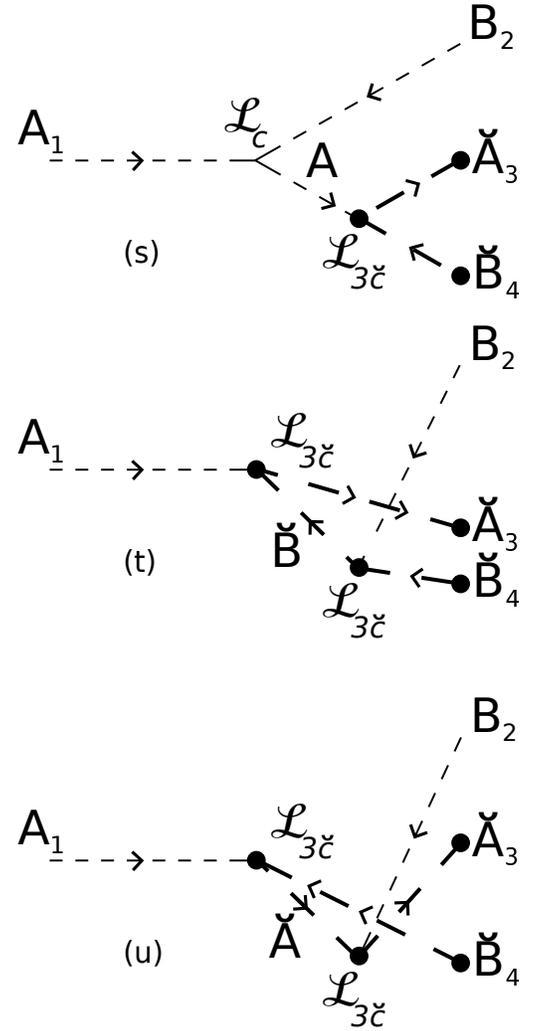}
\caption{\label{fig:epsart} The 3 diagrams for the cascade $A_1 \rightarrow B_2 \breve{A}_3 \breve{B}_4 $ by emission of a particle-antiparticle pair of scalar paraparticles $\breve{A}_3 \breve{B}_4$.}
\end{figure}


\end{document}